\documentclass[12pt]{article}
\usepackage{amssymb}
\usepackage{epsfig,color}
\usepackage{pstricks,graphicx,epsfig,color,amssymb,amsmath,amscd}
\usepackage{cite}
\usepackage[]{graphicx}
\usepackage{makeidx}
\usepackage{amsmath}
\usepackage{nccmath}
\usepackage{setspace}

\usepackage[]{caption}
\renewcommand\rho{\varrho}

\newcommand{\be}{\begin{eqnarray}}
	\newcommand{\ee}{\end{eqnarray}}

\begin{document}
\begin{titlepage}
\title{Mixing of scalar and tensor metric perturbations}
\author{A. D. Dolgov$^{a,b}$, L. A. Panasenko$^{a}$}

\maketitle
\begin{center}
$^a${Department of Physics, Novosibirsk State University, \\Pirogova 2, Novosibirsk 630090, Russia}\\
$^b${Bogolyubov Laboratory of Theoretical Physics, JINR, Dubna,  141980, Russia 
}
\end{center}

\begin{abstract}

Metric perturbations in General Relativity are usually separated into three distinct classes:
 scalar, vector, and tensor. In many cases  these modes are 
 separable, i.e. they satisfy independent
 equations of motion for each mode.
 However, in the present paper we argue
 that in many cases  tensor and scalar modes are not separable,  no matter what gauge 
 conditions are chosen. 
 The propagation of any of these mode depends on the other. A realistic example 
 providing such mixing is presented

\end{abstract}
\thispagestyle{empty}
\end{titlepage}

\section{Introduction \label{s-intro}}
In textbooks \cite{LL-2,mtw,Mukha,Magg,SW,GR-2} and a number of works (for example see ref.~\cite{lifshitz-pert,lif-khal}) metric perturbations are divided into three sections: scalar, vector, and tensor ones. 
In many  cases considered in the literature
these modes are separable, i. e. each mode satisfies a 
separate equation of motion. 
However, in the present paper we show that there may exist realistic cases when tensor and scalar modes are not separable independently on gauge conditions and thus
have an impact to each other propagation.
We start from the expansion of the full metric  $\overline{g}_{\mu\nu}$ over background metric 
${g}_{\mu\nu}$ up to the first order in metric perturbations $h_{\mu\nu}$, where the trace of
the perturbations is defined as $h=h^{\alpha}_{\alpha}$:
\be
&\overline{g}_{\mu\nu}=g_{\mu\nu}+h_{\mu\nu},\\
&\overline{g}^{\mu\nu}=g^{\mu\nu}-h^{\mu\nu},\\
&\overline{g}_{\mu\alpha}\overline{g}^{\mu\beta}=g_{\mu\alpha}g^{\mu\beta}=\delta^{\beta}_{\alpha},\\
&|h| \ll 1,
\ee
The full Einstein equations can be written as:
\be
\overline G_{\mu\nu} &=& 8\pi G \overline T_{\mu\nu},  \label{bar-G-0}\\
\overline G_{\mu\nu} &=&\overline R_{\mu\nu}-\frac{1}{2}\overline g_{\mu\nu}\overline R
\label{G-of-R} ,
\ee 
which are expanded up to the first order in perturbations as:
\be
&\overline{G}_{\mu\nu}=G_{\mu\nu}+G^{(1)}_{\mu\nu},
\label{bar-G}\\
&\overline{T}_{\mu\nu}=T_{\mu\nu}+T^{(1)}_{\mu\nu}
\label{bar-T},
\ee
where the same notations as above
are introduced: overline means a full quantity, upper
index $(1)$ means the first order perturbations, 
the quantities without this index and without bar are the  background ones.

\section{Raising and lowering indices in Einstein equations}

In the previous work we naively assumed that if we lower index in 
{the perturbed energy-momentum tensor (EMT),}
we have to add a new term, proportional to the background {EMT}
multiplied by metric perturbation, i. e.
\be
T^{(1)}_{\mu\nu}=g_{\mu\nu}T^{\mu(1)}_{\nu}+h_{\mu\nu}T^{\mu}_{\nu}.
\ee
{However this is not so.}.
In order to show {that} let us start from the full Einstein {equations} 
{with} mixed indices, because {the} physical EMT is {properly defined} 
with mixed indices, {{$T^{\mu}_{\nu}$}. We 
expand {the} Einstein and EMT {tensors} up to the first perturbation {order as:}
	\be
	G^{\mu}_{\nu}+G^{\mu (1)}_{\nu}=T^{\mu}_{\nu}+T^{\mu (1)}_{\nu},
	\ee
	and lower one index using the full metric
	\be
	G_{\mu\nu}+h_{\mu\alpha}G^{\alpha}_{\nu}+g_{\mu\alpha}G^{\alpha (1)}_{\nu}=T_{\mu\nu}+h_{\mu\alpha}T^{\alpha}_{\nu}+g_{\mu\alpha}T^{\alpha (1)}_{\nu}.
	\ee
	The first {two terms in the l.h.s. and r.h.s.  cancel each other,}
	 because of {the} Einstein equations}  for the background quantities: 
	\be
	&G_{\mu\nu}=T_{\mu\nu},\\
	&h_{\mu\alpha}G^{\alpha}_{\nu}=h_{\mu\alpha}T^{\alpha}_{\nu}.
	\ee
	The resulting equation for the first perturbation order has the form
	\be
	g_{\mu\alpha}G^{\alpha (1)}_{\nu}=g_{\mu\alpha}T^{\alpha (1)}_{\nu}.
	\ee
	Thus we conclude that the first order corrections to the equations of motion with two lower indices are 
	obtained from the exact equations with mixed indices
	by acting of the background metric tensor, but not the full metric tensor, 
	on the first order quantities with mixed indices.
		For example, there is a simple relation
	\be
	T^{(1)}_{\mu\nu}=g_{\mu\nu}T^{\mu(1)}_{\nu}.
	\ee
	
	\section{Helicity expansion and choice of gauge}
	
	To remind the helicity expansion formalism for the metric perturbation $h_{\mu\nu}$  let us write the conventional expansion form, where four scalars, $A, B, E, F$, two vectors $C_i, G_i$, and 
	one tensor,  {$D_{ij}$}, are introduced:
	\be\label{helexp1}
	&&h_{00}=-E,\\
	&&h_{0i}=a\left(\frac{\partial F}{\partial x^i}+G_i\right),\\
	&&h_{ij}=a^2\left(A\delta_{ij}+\frac{\partial^2 B}{\partial x^i \partial x^j}+\frac{\partial C_i}{\partial x^j}+\frac{\partial C_j}{\partial x^i}+D_{ij}\right).\label{helexp3}
	\ee
	Gauge conditions are imposed on {the} scalars to eliminate the non-physical degrees of freedom. There are the {following} two most popular {gauges}: 
	Newtonian, where $B=0=F$ conditions are imposed, and synchronous, where $E=0=F$ conditions are imposed. The last  gauge has {a disadvantage, namely} a residual calibration freedom, that should be fixed for the each {{concrete problem}.
	
	There are some other types of gauges. One can find them for instance in the textbook \cite{SW}. 
	Let us stress that it is necessary to transform EMT too along with the gauge transformations for metric perturbation.
	
	Another {convenient} gauge is the Lorenz gauge, or the Fock harmonic coordinate conditions:
	\be\label{LorenzGgauge}
	D_{\mu}\psi^{\mu}_{\nu}=0,
	\ee
	where
	\be
	\psi^{\mu}_{\nu}=h^{\mu}_{\nu}-\frac{1}{2}h.
	\ee
	This gauge condition complements six independent Einstein equations and leads to a unique solution. One can read about it {in more detail  in ref.}~\cite{GWinberg} or in our previous paper ref.~\cite{OurWork}.
	
	Further in the text we will impose only the Lorenz gauge condition without gauge fixing for the scalar sector.

	
\section{Equation of motion in arbitrary space-time}

{We start from the exact Einstein equations (\ref{bar-G-0},\ref{G-of-R}):}
\be
\bar R_{\mu \nu} - \frac{1}{2} \bar g_{\mu \nu} \bar R = 
{\overline {T}_{\mu \nu}}.
\label{EinEq-tot}
\ee 
{and consider the  first-order tensor perturbations $h_{\mu \nu}$ over an arbitrary 
background metric $g_{\mu \nu}$ }{and expand, as above, guided by eqs.(\ref{bar-G},\ref{bar-T})},  the total Ricci tensor and energy-
momentum tensor as
\be
\bar R_{\mu \nu} = R_{\mu \nu} + R_{\mu \nu}^{(1)}, \ \ \ \ 
{\overline {T}_{\mu \nu}} =  {T}_{\mu \nu} +  {T}_{\mu \nu}^{(1)},  
\label{Rc-EMT-tot}
\ee
 assuming that all background quantities are taken in the background metric $g_{\mu \nu}$.
 
Our goal is to derive the first-order perturbation equation governing evolution of $h_{\mu \nu}$ without any assumptions about  the form of the background metric. 

For the Ricci scalar we have:
\be 
\bar R = \bar g^{\alpha \beta} \bar R_{\alpha \beta} = 
\left(g^{\alpha \beta} - h^{\alpha \beta}\right)\left(R_{\alpha \beta} + R_{\alpha \beta}^{(1)}\right) = 
g^{\alpha \beta}R_{\alpha \beta} - h^{\alpha \beta}R_{\alpha \beta} + g^{\alpha \beta} R_{\alpha \beta}^{(1)}\,.
\label{bar-R-1}
\ee
{ Let us note, that the first order correction to the curvature scalar is not simply obtained from the first order correction to the
Ricci tensor through the contraction of indices by the background metric, but contains an extra term:
\be 
R^{(1)} = g^{\alpha \beta} R_{\alpha \beta}^{(1)} - h^{\alpha \beta}R_{\alpha \beta}.
\label{R1}
\ee} 
Following the book by Landau and Lifshitz \cite{LL-2} we express  the perturbation of the Ricci tensor, 
$R_{\mu \nu}^{(1)}$, via 
metric perturbations, $h_{\mu \nu}$, as
\be
R_{\mu \nu}^{(1)} = \frac{1}{2}\left(D_{\alpha}D_{\nu} h^{\alpha}_{\mu} +
 D_{\alpha}D_{\mu} h^{\alpha}_{\nu} - D_{\alpha}D^{\alpha}h_{\mu \nu} - D_{\mu}D_{\nu}h \right), 
 \label{Ricci-1}
\ee
where covariant derivatives $D_\beta$ are taken with respect to the background metric  $g_{\mu\nu}$ and
$D^{\alpha} = g^{\alpha \beta } D_{\beta}$.
Here and below we use the background metric, $g^{\mu \nu}$ and $g_{\mu \nu}$, to move indices up and down. 

{According to Eq.~\eqref{LorenzGgauge}  
$h^\mu_\nu$ satisfies the condition: 
\be 
D_{\mu} h^{\mu}_{\nu} = \frac{1}{2} \partial_\nu h. 
\label{h-cond}
\ee
}
Using the commutation rules of covariant derivatives we arrive to the result:
\be 
R_{\mu \nu}^{(1)} = - \frac{1}{2}D_{\alpha}D^{\alpha} h_{\mu \nu} +
h^{\alpha \beta} R_{\alpha \mu \nu \beta} + 
 \frac{1}{2}\left(h_{\alpha \mu} R^{\alpha}_{\nu} + h_{\alpha \nu} R^{\alpha}_{\mu}\right).
 \label{Rmn-1}
\ee

Substituting Eqs.~(\ref{Rmn-1}) and (\ref{bar-R-1}) into Eq.~(\ref{EinEq-tot}) and keeping only the first-order quantities
we obtain the following equation for tensor perturbations of the metric:
\be \label{EOM}
&D^2 h_{\mu\nu}-2h^{\alpha\beta}R_{\alpha\mu\nu\beta}-\left(h_{\alpha\mu}R^{\alpha}_{\nu}+h_{\alpha\nu}R^{\alpha}_{\mu}\right)+h_{\mu\nu}R-g_{\mu\nu}\left(h^{\alpha\beta}R_{\alpha\beta}+\frac{1}{2}D^2 h\right)\nonumber\\
&=-2 (8\pi G)T^{(1)}_{\mu\nu}.
\ee
	
\section{Mixing of scalar and tensor modes}

	Let us stress that, {as it follows from  eq.~(\ref{EOM})},  in general case 
	{of an arbitrary form of the Ricci tensor there may be a mixing  of the scalar and tensor modes of metric perturbations.} 
	Indeed, taking trace from eq.(\ref{EOM}) we obtain 
	\be\label{EOMtrace}
	\partial^2h+4h^{\alpha\beta} R_{\alpha\beta}- hR=16\pi G T^{\alpha (1)}_{\alpha} .
	\ee
	{From} here it is clear that convolution in the second term can include both scalar and tensor  
	parts. In addition, it is worth to pay attention to the {right hand side of the  equation, 
	that contains the trace of }	the source. As it will be shown further, for the {problem of the} graviton to photon conversion in external magnetic field, this trace contains a convolution of background electromagnetic tensor $F^{\mu\nu}$ and perturbation tensor $h_{\mu\nu}$, what consequently leads again to the mixing of scalar and tensor modes in this special {case}.
	
	{Thus we conclude, that tensor-scalar mixing could generally originate from:}\\
	1.  Possible explicit mixing in the l.h.s. of eq.~(\ref{EOMtrace}) that can be provided by the 
	form of the Ricci tensor,\\
	2. Possible implicit mixing  in the r.h.s. of eq.~(\ref{EOMtrace}) that can be provided by the form of the source $T^{\mu (1)}_{\nu}$.\\
	The last important note in this section is about gauge independence of the result. We derived  {eq.~(\ref{EOM})} without introducing any gauge condition, besides the Fock harmonic calibration $D_{\mu}h^{\mu}_{\nu}=0$. {In other words we had not impose  any condition} on the scalar sector, and we think that neither synchronous nor Newtonian gauge conditions can {eliminate the scalar-tensor mixing}
	because the mixing originates from the form of the Ricci tensor and/or from the form of the source. 
	
{Maybe} the Ricci tensor form, providing the mixing, should be non-trivial or even exotic. 
We will {look for some possibilities} in the future work. But what we can say {for sure}, that {there is at} least one physically realistic case for the {form of the} energy-momentum tensor perturbation  to {create a}  mixing of scalar and tensor modes in the first perturbation order.

	\section{Graviton to photon conversion in FLRW space-time and external magnetic field}
	
	There are several works about graviton to photon transition in external magnetic field ref.~\cite{g-to-gam1,g-to-gam2,g-to-gam3,g-to-gam4,g-to-gam5,g-to-gam6,g-to-gam7,g-to-gam8,g-to-gam9} but only in Minkowski flat space-time and there no consideration of scalar and tensor mode mixing.
	
	In this section we consider  {the 3D flat Friedmann-LeMaitre-Robertson-Walker (FLRW) metric with the  line element:
	\be
	ds^2 = dt^2 - a^2(t) \delta_{ij} dx^i dx^j
	\label{ds-2}
	\ee
	}
	Remind that in {the FLRW space-time}
	\be\label{Ricci}
	&R_{00}=-3\frac{\ddot{a}}{a},\\
	&R_{ij}=-g_{ij}\left(\frac{\ddot{a}}{a}+2H^2\right).\\
	&R=-6\left(\frac{\ddot{a}}{a}+H^2\right).
	\ee
	{The trace of metric perturbations can be written} in the following form:
	\be\label{hii}
	h=g^{\mu\nu}h_{\mu\nu}=h_{00}-\frac{h_{xx}+h_{yy}+h_{zz}}{a^2}\equiv h_{00}+h^i_i,
	\ee
	{where  $h^i_i=-\left({h_{xx}+h_{yy}+h_{zz}}\right)/{a^2}$.}
	
	Using equations (\ref{Ricci}-\ref{hii}), we can rewrite eq.(\ref{EOMtrace}) as {follows:}
	\be\label{EOMtraceFLRW}
	\partial^2 h-12\frac{\ddot{a}}{a}h_{00}-4\left(\frac{\ddot{a}}{a}+2H^2\right)h^i_i+
	6\left(\frac{\ddot{a}}{a}+H^2\right)h=16\pi G T^{\alpha(1)}_{\alpha} .
	\ee

	Now our goal is to find an explicit form for the right side of  {{eq.(\ref{EOMtraceFLRW}).}
	From the Maxwell and {Heisenberg-Euler~\cite{HE}} actions we can find energy-momentum tensor corrections. {The corresponding} actions have the forms
	\be
	&{\cal A}{_{Max}} = -\frac{1}{4}\int d^4x  \sqrt{- \bar g} \left( \bar F^2 + \bar A_\mu \bar J^\mu \right)
	\label{A-Max},\\
	&{\cal A}_{HE} =  \int d^4x \sqrt{-g} \, C_0\left[(\overline{F}_{\mu\nu} \overline{F}^{\mu\nu})^2+
	\frac{7}{4}(\overline{{\tilde F}}^{\mu\nu} \overline{F}_{\mu\nu})^2 \right] ,
	\label{A-HE}
	\ee
	{where $C_0 = \alpha^2/(90 m_e^4)$ and $\alpha = 1/137$ is the fine structure constant.
	At high temperatures $C$, $\alpha$, and $m_e$ change with $T$.
	The dual Maxwell tensor is defined as
	\be 
	\tilde F_{\alpha\beta} = \frac{\sqrt{-g}}{2}\,  \epsilon_{\alpha\beta\mu\nu} F^{\mu\nu}, \,\,\,
	\tilde F^{\alpha\beta} = \frac{1}{2 \sqrt{-g}} \epsilon^{\alpha\beta\mu\nu} F_{\mu\nu}, 
	\label{tilde-F-c}
	\ee
	because the tensor quantity is $\sqrt{-g}\,  \epsilon_{\alpha\beta\mu\nu} $ but not just
	$\epsilon_{\alpha\beta\mu\nu}$, see  e.g. chapter 83 from  textbook~\cite{LL-2}.}
	
	Let us briefly remind that {the} Heisenberg-Euler effective Lagrangian
	 describes quartic self-interaction of electromagnetic field. It is induced  
	by the loop of virtual electrons with four external electromagnetic legs. In the weak field limit,
	and low energies, much smaller than the electron mass, $m_e$, the corresponding action has the form of eq. (\ref{A-HE}). In what follows we apply this action to photon propagation in external magnetic field $B$ {in} the weak field limit: {$B\ll m_e^2$.}
	
	{The first perturbation order of the EMT has the form:}
	\be\label{EMTMax}
	T^{Maxwell(1)}_{\mu\nu}=\frac{1}{2}g_{\mu\nu}\left[Ff-FFh\right]+h_{\mu\sigma}F^{\sigma\alpha}F_{\nu \alpha}+h^{\alpha\sigma}F_{\mu\alpha}F_{\nu\sigma}-f_{\mu\alpha}F_{\nu}^{.\,\alpha}-F_{\mu\alpha}f_{\nu}^{.\,\alpha},
	\ee
	\be\label{EMTHE}
	&&T_{\mu\nu}^{HE\,(1)} = C(T) \left[ - h_{\mu\nu}  (F^2)^2 
	- 4 F^2 g_{\mu\nu} \left( Ff - FFh \right) +  
	8  F^2  (f_{\nu\lambda} F_\mu^{.\,\lambda}
	+f_{\mu\lambda} F_\nu^{.\,\lambda})
	\nonumber \right.\\
	&&+16 F_{\mu\lambda}  F_\nu^{.\,\lambda} (Ff)
	\left.
	- 8h^{\lambda\sigma} F_{\mu\lambda}  F_{\nu\sigma} F^2
	-16 F_{\mu\lambda} F_\nu^{.\,\lambda} (FFh)
	\right].
	\ee
	where $Ff=F^{\alpha\beta}f_{\alpha\beta},\,\,\,
	FFh=h^{\alpha}_{\sigma}F_{\alpha\beta}F^{\sigma\beta}$.
	
	EMT correction originated from {the} Maxwell action is traceless, but {those} originated from {the} Hisenberg-Euler action is not. Indeed, we have
	\be
	T_{\alpha}^{\alpha\,HE\,1} = C(T) \left[-hF^4+16F^2 (Ff) - 8 F^2 (FFh)\right]
	\label{trace-T-HE1}
	\ee
	
	Let us write eq.(\ref{EOMtraceFLRW}) with the r.h.s. {explicitly:} 
	\be\label{EMTcorrectiontrace}
	&\partial^2 h-12\frac{\ddot{a}}{a}h_{00}-4\left(\frac{\ddot{a}}{a}+2H^2\right)h^i_i+6\left(\frac{\ddot{a}}{a}+H^2\right)h=\nonumber\\
	&=16\pi GCF^2\left[-hF^2+16(Ff) - 8 {(F_{\alpha\beta}F^{\alpha\sigma}h^{\beta}_{\sigma})}\right].
	\ee
	{Let us choose the direction of the external magnetic field along the $x$ axis.
	In such coordinates the non-zero components of electromagnetic tensor are the following:}
	\be
	&F^y_{.\,z}&=-F^z_{.\,y}=B_x.\\
	&F^{yz}&=-F^{zy}=-\frac{B_x}{a^2}.\\
	&F_{yz}&=-F_{zy}=-B_x a^2.
	\ee
	{For what follows we need the expressions:}:
		\begin{gather}
	F^2=2B^2,\\
	Ff=F^{.\,\beta}_{\alpha}f^{\alpha}_{.\,\beta}=F^{.\,z}_y f^y_{.\,z}+F^{.\,y}_z f^z_{.\,y}=B\left(f^y_{.\,z}-f^z_{.\,y}\right)=2Bf^y_{.\,z}\label{Ff},\\
	FFh=h^{\alpha}_{\sigma}F_{\beta\alpha}F^{\beta\sigma}=B^2\left(h^y_y+h^z_z\right)\label{FFh}.
	\end{gather}

	{After substituting them into eq.(\ref{EMTcorrectiontrace}) we obtain}
	\be\label{EMTcorrectiontraceExplicit}
	&\partial^2 h-12\frac{\ddot{a}}{a}h_{00}-4\left(\frac{\ddot{a}}{a}+2H^2\right)h^i_i+
	6\left(\frac{\ddot{a}}{a}+H^2\right)h=\nonumber\\
	&=32\pi GCB^3\left[-hB+16f^y_{.\,\,z} - 4 {B\left(h^y_y+h^z_z\right)}\right].
	\ee
	Diagonal elements of metric perturbations can be written as a sum of scalar and tensor quantities in {terms of the helicity decomposition} (see eq.(\ref{helexp1}-\ref{helexp3}))
	{\be
	h^y_y=2\Psi \delta^y_y +D^y_y,\\
	h^z_z=2\Psi \delta^z_z +D^z_z.
	\ee}
	{Here we finally use {the Newtonian gauge} in the scalar sector and impose $E\equiv 2\Phi$, $A\equiv-2\Psi$ to be more specific, but  the mixing in eq.(\ref{EMTcorrectiontraceExplicit}) does not depend on the calibration because {anyway we have the tensor} part in the r.h.s.}
	
	Finally we {obtain} the equation including both scalar $\Phi, \Psi$ and tensor modes $D_{ij}$:	
	\be
	&\left(\partial^2+32\pi G CB^4\right) \left(\Phi+3\Psi\right)+6\left(\Psi-\Phi\right)\left(\frac{\ddot{a}}{a}-H^2\right)=\nonumber\\
	&=16\pi GCB^3\left[16f^y_{.\,\,z} - 4 {B\left(4\Psi+D^y_y+D^z_z\right)}\right],
	\ee
	where $h=2\Phi+6\Psi$.
	
	Moreover, there can be non-explicit mixing of modes through $f^y_{.\,\,z}$, but to show that we have to write derivation of electromagnetic field equation of motion, what can make the article too {long}. {Complete solution for the  graviton-photon system  will be considered elsewhere}.
	
	\section{Conclusion}
	In the present paper we derived the equation for metric perturbations {to study} whether it is possible to separate scalar and tensor modes or they mix with each other. In the {latter case one cannot separate the equations of motion for these two modes. As a consequence in some special cases there can be a mutual influence of the tensor mode on the scalar one and vice versa, i.e. one initial mode can excite another one in the process of propagation.}
	
	{We {have also given} an example where such mixing is present, namely  the process of the} graviton to photon conversion in the external magnetic field in FLRW space-time. Such conditions {could exist for the gravitational wave transition to photon in} magnetized cloud of matter, or for primary metric perturbations, generated by inflation (for more detail see ref.~\cite{Grishchuk,StarGW,VerRubSazh,parker1,parker2,aas-pert,ADL-infl}), in the primeval plasma {with presence} of cosmological magnetic fields. Therefore, the mixing phenomenon can significantly change primary metric perturbation spectra both in tensor and  scalar sectors }and can have {observable consequences for the present day universe.} We will devote our future investigation to this issue.
	
\section*{ Acknoledgement} 
The work was supported by RSF grant № 23-42-00066.

\end{document}